\documentclass[12pt]{article}
\usepackage{epsfig}
\usepackage{ulem}
\usepackage{graphicx}
\usepackage{cite}
\usepackage{amsfonts}
\usepackage{amssymb}
\usepackage{bm}
\usepackage{latexsym}
\usepackage{amsmath}
\def\ee{\end{equation}}
\def\ba{\begin{eqnarray}}
\def\ea{\end{eqnarray}}

\def\bq{\begin{quote}}
\def\eq{\end{quote}}

 at 10truept

\newcommand{\beq}{\begin{equation}}
\newcommand{\eeq}{\end{equation}}
\newcommand{\beqa}{\begin{eqnarray}}
\newcommand{\eeqa}{\end{eqnarray}}
\newcommand{\bea}{\begin{eqnarray}}
\newcommand{\eea}{\end{eqnarray}}
\newcommand{\p}{\partial}

\newcommand{\lle}{\left<}
\newcommand{\rgr}{\right>}
\newcommand{\lb}{\left|}
\newcommand{\rb}{\right|}

\newcommand{\Torder}{\mathrm{T}}
\newcommand{\vect}[1]{\bm{\mathrm{{#1}}}}
\newcommand{\hf}{\frac{1}{2}}
\newcommand{\khat}{{\hat q}}

\def\ltap{\ \raise.3ex\hbox{$<$\kern-.75em\lower1ex\hbox{$\sim$}}\ }
\def\gtap{\ \raise.3ex\hbox{$>$\kern-.75em\lower1ex\hbox{$\sim$}}\ }
\def\gl{\ \raise.5ex\hbox{$>$}\kern-.8em\lower.5ex\hbox{$<$}\ }
\def\roughly#1{\raise.3ex\hbox{$#1$\kern-.75em\lower1ex\hbox{$\sim$}}}
\newcommand\vx{{\vect x}}
\newcommand\vxp{{\vect x}'}
\newcommand\vy{{\vect y}}

\newcommand\vk{{\vect k}}
\newcommand\vkp{{\vect k}'}

\newcommand\dslq{{\overline {dq}}}  %\textrm{\sout{\it dq}}

\begin{document}

\thispagestyle{empty}
\begin{titlepage}
\nopagebreak

\title{%\bf\sc\LARGE 
Cosmological diagrammatic rules} 
%\begin{center}
%\end{center}

\vfill
\author{Steven B. Giddings$^{a}$\footnote{giddings@physics.ucsb.edu}\ \  and Martin S. Sloth$^{b}$\footnote{sloth@cern.ch}}
\date{ }

%\end{center}

\maketitle

\vskip 0.5cm

%\begin{center}
{\it  $^{a}$ Department of Physics, 
University of California, Santa Barbara, CA 93106}
\vskip 0.5cm
{\it  $^{b}$ CERN, Physics Department, Theory Unit, CH-1211 Geneva 23, Switzerland}

%\end{center}
\vfill
\begin{abstract}
A simple set of diagrammatic rules is formulated for perturbative evaluation of ``in-in" correlators, as is needed in cosmology and other nonequilibrium problems.  These rules are both intuitive, and efficient for calculational purposes.

 \end{abstract}
 \vfill
 
\noindent
CERN-PH-TH/2010-108\hfill \\  
\vfil
\end{titlepage}

%\paragraph{The "in-in" formalism.}

When calculating quantities relevant for cosmological evolution, one needs an efficient means to calculate operator expectation values, or more complicated expressions, in the ``in-in," or Schwinger-Keldysh, formalism.\footnote{For reviews, see for example \cite{Chou:1984es,Calzetta:1986ey,Weinberg:2005vy,Chen:2010xk}.}  Various methods or rules have been derived to do this, in particular the closed time path formalism, and the rules outlined in \cite{Weinberg:2005vy} and \cite{Musso:2006pt,vanderMeulen:2007ah,Petri:2008ig}.  This note formuates a refined set of such diagrammatic rules that seems both intuitive, and  is efficient for calculational purposes. For example, this reduces the calculational complexity common in uses of the closed time path formalism.

As a simple and concrete example, take $\phi^3$ theory, 
\beq\label{phithree}
{\cal L}=-\frac{1}{2}\p_{\mu}\phi\p^{\mu}\phi -\frac{g}{3!}\phi^3
\eeq
in a fixed background Robertson-Walker metric,
\beq
ds^2 = -dt^2+a^2(t)ds_3^2\ ,
\eeq
where $ds_3^2$ is the metric of a homogeneous space.
It is useful to work with conformal time, $\eta$, instead of physical time, $t$, which is defined by
\beq
a(\eta)d\eta = dt~.
\eeq

In the in-in formalism the expectation value of any operator $\mathcal{O}$
(evaluated at time $\eta_0$, and up to vacuum normalization)
is given by
\begin{equation}
    \label{exp1}
    \lle \Omega \rb \mathcal{O}(\eta_0) \lb \Omega \rgr =\lle
    0\rb
    {\bar \Torder}\left(e^{i\int_{-\infty}^{\eta_0}d\eta H_I}\right) \mathcal{O}(\eta_0) T\left(e^{-i\int_{-\infty}^{\eta_0} d\eta H_I}\right)\lb
    0 \rgr
\end{equation}
where $\lb \Omega \rgr$ is the vacuum of the interacting theory,
$\lb 0\rgr$ is the
vacuum of the free theory,
$\Torder$ and $\bar \Torder$ are time ordering and anti-ordering operators, respectively, and $H_I$ is the
interaction hamiltonian in time $\eta$.  The expectation value can be evaluated by expanding the exponential, and contracting fields, as in the usual Wick analysis of in-out amplitudes in flat-space field theory.

Concretely, consider the operator ${\cal O}= \phi(\vx,\eta_0)\phi(\vxp,\eta_0)$.  The correction to the corresponding propagator to second order in $g$ follows from the second-order terms in the expansion of the exponentials.  There are two kinds of terms.  The first is
\beq\label{Aamp}
A(\eta_0,\vx,\vxp)=\langle0|   \phi(\vx,\eta_0)\phi(\vxp,\eta_0)T \frac{1}{2!} \int dy \frac{-ig}{3!} \phi^3(y) \int dy' \frac{-ig}{3!} \phi^3(y')|0\rangle\ ,
\eeq
with a corresponding term also from the left exponential.  Here we use notation $y=(\eta,\vy)$, $dy=a^4 d^3{\vect y}d\eta$, and the $\eta$ integrals range up to $\eta_0$.  The second kind of term comes from the linear expansion in each of the exponentials:
\beq\label{Bamp}
\langle0|{\bar \Torder}\int dy \frac{ig}{3!} \phi^3(y)\,   \phi(\vx,\eta_0)\phi(\vxp,\eta_0)\, T   \int dy' \frac{-ig}{3!} \phi^3(y')|0\rangle\ .
\eeq

Two kinds of propagator enter the corresponding expressions, the Wightman propagator,
\beq
W(x,x') =\langle0|\phi(x)\phi(x')|0\rangle\ ,
\eeq
and the Feynman propagator,
\beq
G(x,x') =\langle0|T\phi(x)\phi(x')|0\rangle\ .
\eeq
Both expressions may be evaluated directly in terms of corresponding Wick contractions.  

For (\ref{Aamp}), we find 
\beq
A(\eta_0,\vx,\vxp)= \hf \int dy (-ig) \int dy' (-ig) W(x,y) G(y,y')^2 W(x',y')\ .
\eeq
We could also directly find the term from expanding the left exponential, but observe that it simply gives the complex conjugate expression, with a combined contribution $2{\rm Re} A$.

Likewise, the contractions from (\ref{Bamp}) provide two separate terms.  The first is 
\beq
B(\eta_0,\vx,\vxp) = \hf \int dy (ig) \int dy' (-ig) W(y,x) W(x',y') W^2(y,y')\ .
\eeq
The second is again the complex conjugate expression.

%\paragraph{The diagrammatic rules.}
The expressions above are given by the following rules:

\begin{enumerate}

\item Draw a horizontal dotted line, corresponding to $\eta_0$, and place the external points of the correlator on this line.

\item  At a given order, enumerate all placements of vertices either above or below the $\eta_0$ line, modulo reflections about this line.  Then, draw all diagrams connecting these vertices with propagator lines, again modulo reflection.

\item Each propagator line crossing or ending on the dotted line gives a Wightman propagator, whose leftmost/rightmost time argument corresponds to the uppermost/lowermost vertex.
Each propagator line below the dotted line gives a Feynman propagator, and each line above the dotted line gives the complex conjugate or time-reversed Feynman propagator. 

\item Each vertex below the dotted line gives a $V=-ig$ together with an integral, and each above gives an $V^\dagger =ig$ and an integral.

\item Divide by the usual Feynman symmetry factors, where present.

\item Once the resulting diagrams are calculated, take twice their real part.

\end{enumerate}

The corresponding diagrams for the amplitudes $A(\eta_0,\vx,\vxp)$ and $B(\eta_0,\vx,\vxp)$ are shown in fig.~1.  The $1/2$ in each corresponding expression is a symmetry factor.

\begin{figure}[!hbtp] \begin{center}%\label{fig1}
\includegraphics[width=16cm]{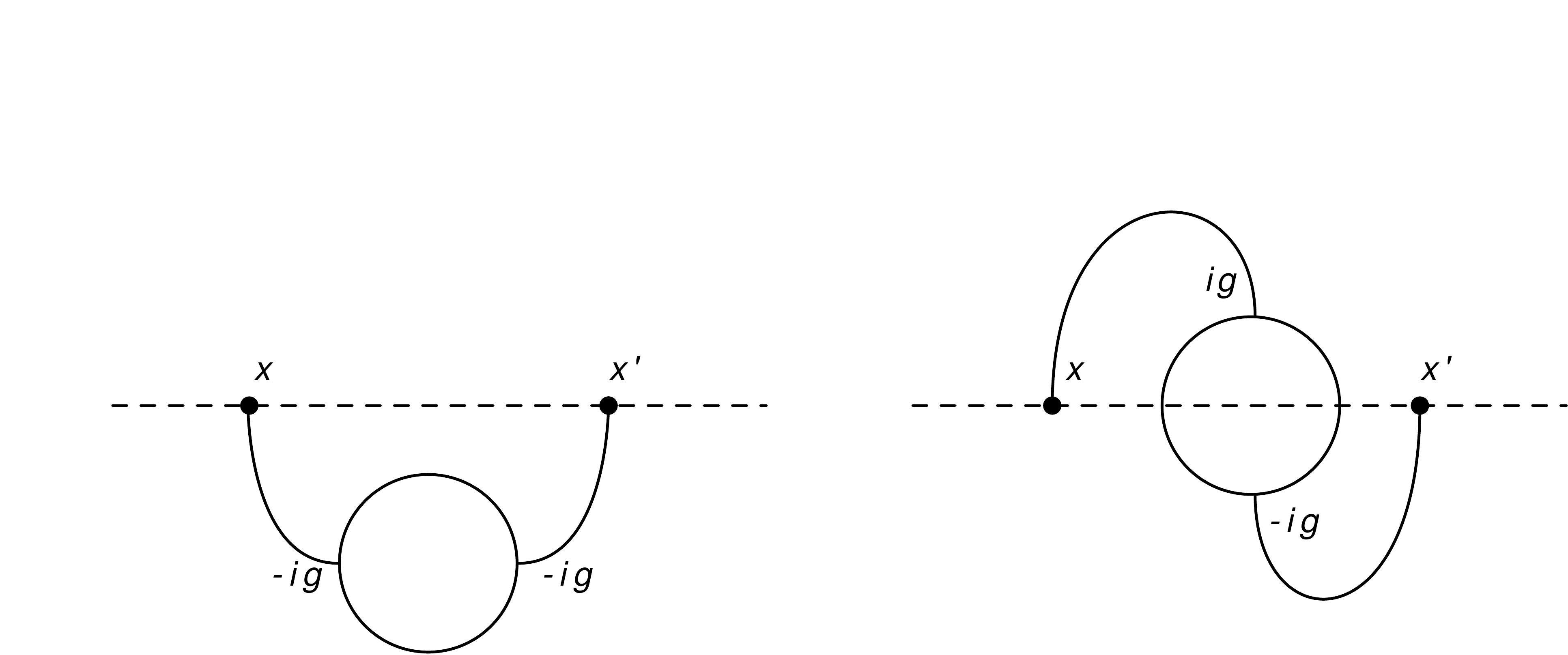}
\end{center}
\caption{The bubble diagrams corresponding to expressions (\ref{Aamp}), (\ref{Bamp}).}
\end{figure}

If one works in a cosmological spacetime with a flat slicing, so that linear spatial momentum is conserved, these rules are easily reformulated working in momentum space (but retaining time $\eta$ as parameter).  Specifically, in rule (4) one uses the momentum-space version of the propagator, and we replace

\medskip
\noindent5.$\rightarrow5.' $ Vertices below/above the line are accompanied by $V$ or $V^\dagger$, respectively; conserve momentum at each vertex and include an overall momentum-conserving delta function, integrate over all internal momenta, and integrate over the time coordinate of each vertex.  
\medskip

Clearly these rules have a trivial generalization to theories with more fields and more complicated vertices.

\paragraph{Examples.}

These rules appear to offer modest streamlining of existing calculations.  For example, via these rules one immediately writes down (\ref{Aamp}) and (\ref{Bamp}), or the momentum-space expressions, 
\beq\label{Amom}
2{\rm Re}A(\eta_0,\vk,\vkp)=\ -g^2 (2\pi)^3\delta^3(\vk+\vkp) {\rm Re}\int a^4 d\eta   a^4 d\eta'  \int\dslq W_k(\eta_0,\eta) G_{q}(\eta,\eta')G_{|\vect k-\vect q|}(\eta,\eta') W_k(\eta_0,\eta')
\eeq
\beq\label{Bmom}
2{\rm Re}B(\eta_0,\vk,\vkp)=\ g^2 (2\pi)^3\delta^3(\vk+\vkp){\rm Re} \int a^4 d\eta a^4 d\eta'  \int\dslq W_k(\eta,\eta_0) W_{q}(\eta,\eta')W_{|\vect k-\vect q|}(\eta,\eta') W_k(\eta_0,\eta')
\eeq
where a useful shorthand for calculations is  $\dslq = d^3q/(2\pi)^3$.
In the special case of de Sitter space, we then use the Wightman propagator
\beq
 \left<\phi_{\vect k}(\eta)\phi_{\vect k'}(\eta')\right> = (2\pi)^3\delta^3(\vect k+\vect k')W_k(\eta,\eta'),
 \eeq
 with
  \beq\label{Wscalar}
 W_k(\eta,\eta')=U_{k}(\eta) U^*_{k}(\eta')\ .
 \eeq
 and
 \beq
 U_k(\eta)=
    \frac{H}{\sqrt{2k^3}}(1+ik\eta)e^{-ik\eta}\ , 
 \eeq
 and the corresponding Feynman propagator, with
 \beq
G_k(\eta,\eta')=\theta(\eta-\eta')W_k(\eta,\eta') + \theta(\eta'-\eta)W_k(\eta',\eta)\ .
\eeq

We could also consider the same type of diagram, but with graviton-scalar interactions, {\it e.g.}
\beq\label{lthree}
{\cal L}_3 = \frac{a}{2}\gamma_{ij}\p_i\sigma\p_j\sigma\ 
\eeq
and using the graviton propagator in transverse traceless gauge, 
\beq
\langle\gamma_{ij}(\vect k,\eta)\gamma_{kl}(\vect k',\eta')\rangle = (2\pi)^3\delta^3(\vect k+\vect k')2\omega_{ij,kl}(k)W_k(\eta,\eta')
\eeq
with polarization sum
\bea
\omega_{ij,kl}(\vect q)&=& \delta_{ik}\delta_{jl}+ \delta_{il}\delta_{jk}-\delta_{ij}\delta_{kl}\\
&+&\delta_{ij}\khat_k\khat_l + \delta_{kl}\khat_i\khat_j- \delta_{ik}\khat_j\khat_l - \delta_{il}\khat_j\khat_k-\delta_{jk}\khat_i\khat_l-\delta_{jl}\khat_i\khat_k + \khat_i\khat_j\khat_k\khat_l\nonumber
\eea
one immediately reproduces the two one-loop bubble diagrams given in eqs. (3.14) and (3.15) of \cite{Giddings:2010nc}.  Likewise, with external gravitons, one has an immediate derivation of eq. (20) of \cite{Adshead:2009cb}.  In some cases, like the graviton bubble of \cite{Giddings:2010nc}, it is then possible to explicitly perform the $\eta$ integrals to find elementary expressions.

\begin{figure}[!hbtp] \begin{center}%\label{fig2}
\includegraphics[width=15cm]{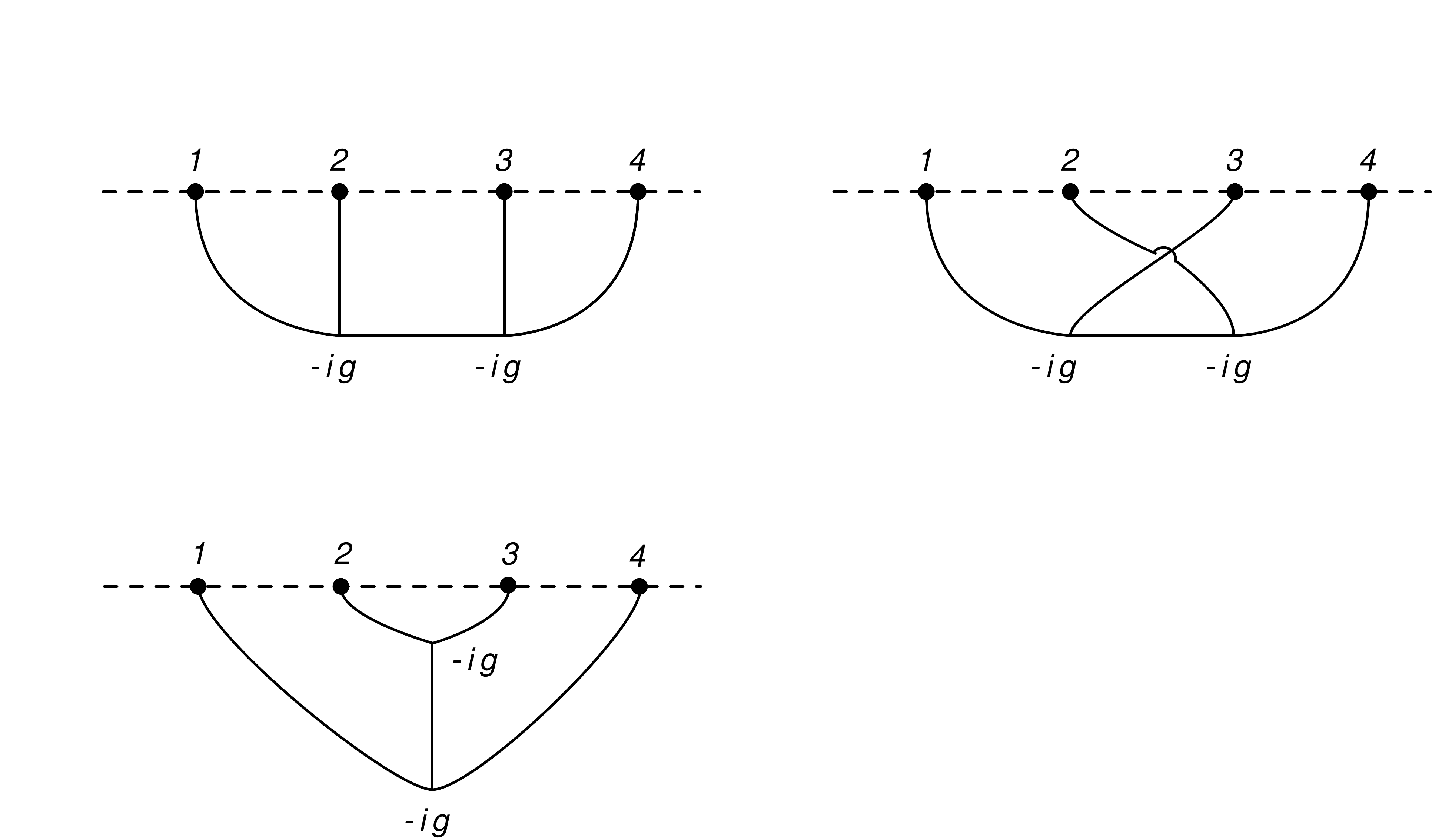}
\end{center}
\caption{Diagrams contributing to the trispectrum, through the second line of (\ref{trispec}). }
\end{figure} 

\begin{figure}[!hbtp] \begin{center}%\label{fig3}
\includegraphics[width=15cm]{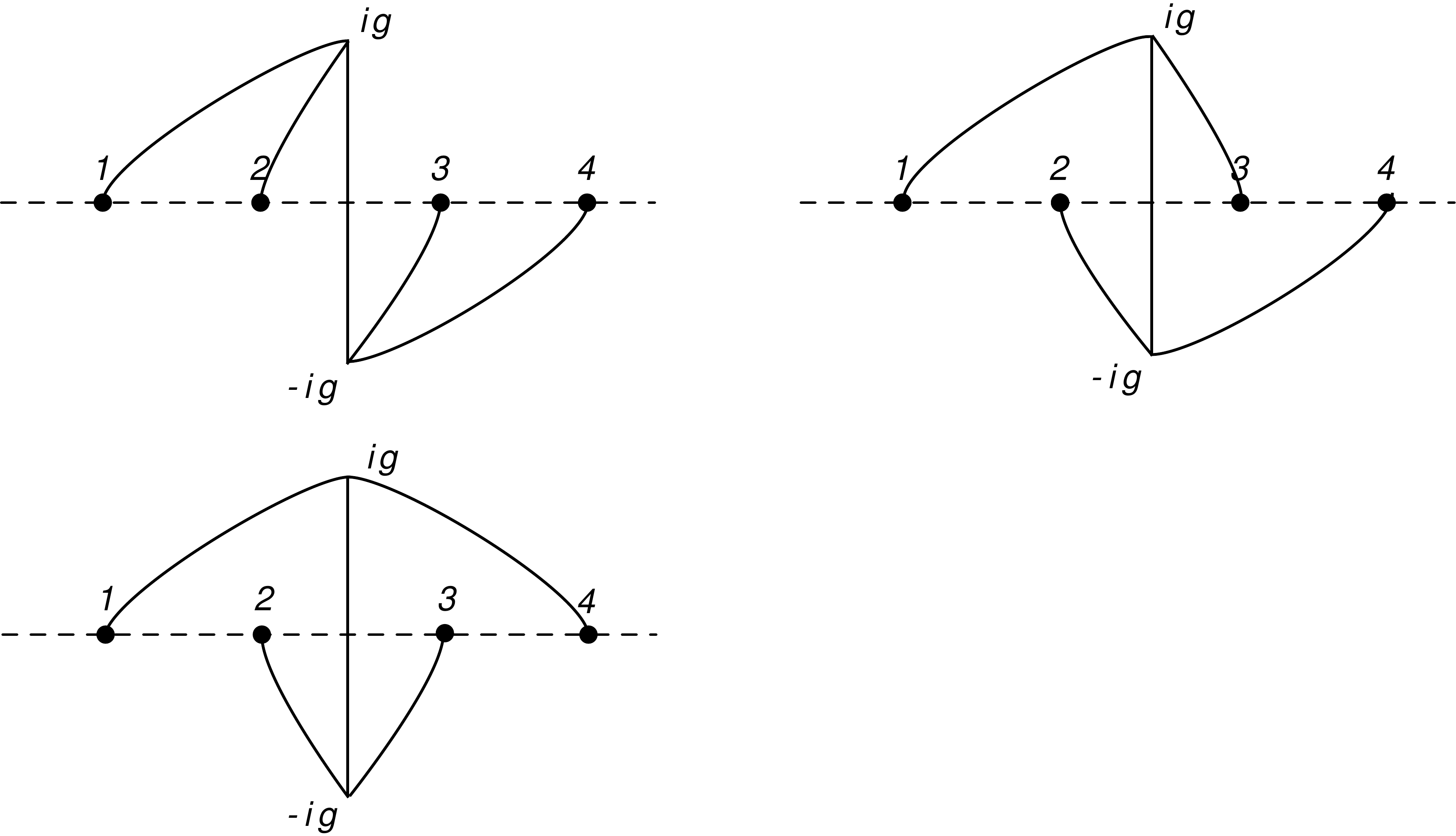}
\end{center}
\caption{Diagrams contributing to the trispectrum, through the third line of (\ref{trispec}).}
\end{figure}

Another example is the tri-spectrum, calculated in \cite{Seery:2008ax}.  There we have the six diagrams of fig.~2 and fig.~3, once reflection symmetry is accounted for.  Then (taking the simpler case of an exchanged scalar, via the interaction (\ref{phithree})), we immediately write down the amplitude 
\bea\label{trispec}
\langle\phi_{\vect k_1}\phi_{\vect k_2}\phi_{\vect k_3}\phi_{\vect k_4}\rangle &=& -g^2(2\pi)^3 \delta^3\left(\sum_i \vect k_i\right)\cdot 2{\rm Re}\Biggl[\int_{-\infty}^{\eta_0} a^4d\eta \int_{-\infty}^{\eta_0} a^4d\eta' \\
&&W_{k_1}(\eta_0,\eta) W_{k_2}(\eta_0,\eta) W_{k_3}(\eta_0,\eta') W_{k_4}(\eta_0,\eta') G_{|\vect k_1+\vect k_2|}(\eta,\eta')\nonumber\\
&- & W_{k_1}(\eta,\eta_0) W_{k_2}(\eta,\eta_0) W_{k_3}(\eta_0,\eta') W_{k_4}(\eta_0,\eta') W_{|\vect k_1+\vect k_2|}(\eta,\eta')
\Biggr]+(2\leftrightarrow3) + (2\leftrightarrow4)\nonumber
\eea
to be compared with (2.20) and subsequent formulas in  \cite{Seery:2008ax}.

Thus, we do find  calculational streamlining, which we expect to be more significant for more complicated diagrams.

%\newpage

\end{document}